\documentclass[twocolumn,showpacs,preprintnumbers,nofootinbib,prd,superscriptaddress,groupedaddress,10pt]{revtex4-1}

\usepackage{graphicx,amssymb,amsmath,amsthm,amsfonts,epsfig,epsf}
\usepackage[linktocpage]{hyperref}
\usepackage[usenames]{color}

\usepackage{aas_macros}
\usepackage{bm}
\usepackage{dcolumn}
\usepackage[latin1]{inputenc}
\usepackage{latexsym}
\usepackage{rotating}
\usepackage{hyperref}
\usepackage{color}
\usepackage{longtable}
\usepackage{enumerate}
\usepackage{tensor}
\usepackage{mathtools}
\usepackage{url}
\def\nn{\nonumber}

\newcommand{\ben}{\begin{enumerate}}
\newcommand{\een}{\end{enumerate}}

\def\be{\begin{equation}}
\def\ee{\end{equation}}
\def\bea{\begin{eqnarray}}
\def\eea{\end{eqnarray}}
\newcommand{\beq}{\begin{eqnarray}}
\newcommand{\eeq}{\end{eqnarray}} 
\newcommand{\ba}{\begin{align}}
\newcommand{\ea}{\end{align}}

\def\nn{\nonumber}

\begin{document}

\title{Ultra-high-energy debris from the collisional Penrose process}

\author{
Emanuele Berti$^{1}$\footnote{Electronic address: eberti@olemiss.edu},
Richard Brito$^{2,3}$\footnote{Electronic address: richard.brito@tecnico.ulisboa.pt},
Vitor Cardoso$^{1,2,3}$\footnote{Electronic address: vitor.cardoso@tecnico.ulisboa.pt}
}
\affiliation{${^1}$ Department of Physics and Astronomy, The University of Mississippi, University, MS 38677, USA}
\affiliation{${^2}$ CENTRA, Departamento de F\'{\i}sica, Instituto Superior T\'ecnico -- IST, Universidade de Lisboa -- UL,
Avenida Rovisco Pais 1, 1049 Lisboa, Portugal}
\affiliation{${^3}$ Perimeter Institute for Theoretical Physics Waterloo, Ontario N2J 2W9, Canada}
%


\begin{abstract}
Soon after the discovery of the Kerr metric, Penrose realized that
superradiance can be exploited to extract energy from black holes.
The original idea (involving the breakup of a single particle) yields
only modest energy gains. A variant of the Penrose process consists of
particle collisions in the ergoregion. The collisional Penrose process
has been explored recently in the context of dark matter searches,
with the conclusion that the ratio $\eta$ between the energy of
post-collision particles detected at infinity and the energy of the
colliding particles should be modest ($\eta \lesssim
1.5$). Schnittman~\cite{Schnittman:2014zsa} has shown that these
studies underestimated the maximum efficiency by about one order of
magnitude (i.e., $\eta \lesssim 15$). 
In this work we show that particle collisions in the vicinity of
rapidly rotating black holes can produce high-energy ejecta and result
in high efficiencies under much more generic conditions.
The astrophysical likelihood of these events deserves
further scrutiny, but our study hints at the tantalizing possibility
that the collisional Penrose process may power gamma rays and
ultra-high-energy cosmic rays.
\end{abstract}


\pacs{04.70.-s,04.70.Bw,04.70.Dy}


\maketitle


\noindent{\bf{\em I. Introduction.}}~It is tempting to say that modern
relativistic astrophysics was born in 1963, when Kerr discovered the
famous solution of Einstein's equations describing rotating black
holes~\cite{Kerr:1963ud} and Schmidt identified the first
quasar~\cite{1963Natur.197.1040S}. The connection between these
theoretical and observational breakthroughs became clear much later,
but astronomers now agree that black holes are the engines behind many
high-energy events in the Universe. Black holes are the most compact
objects in Nature, and therefore particles in their vicinities can
attain relativistic velocities, producing observable phenomena such as
jets and gamma-ray bursts. The mechanisms converting the gravitational
energy stored by black holes into high-energy fluxes of matter and
radiation are a major area of investigation in relativistic
astrophysics, and particle collisions near black holes could have
important implications for dark matter searches (see
e.g.~\cite{Gondolo:1999ef,Fields:2014pia}).

In 1969, Penrose discovered a remarkable way to extract energy from
Kerr black holes~\cite{Penrose:1969}. In the so-called ``ergoregion''
surrounding rotating black holes particles can have negative energies
$(\epsilon_i<0)$ as measured by an observer at infinity. Therefore an
object of energy $\epsilon_0$ can fragment into two pieces, one of
which escapes to infinity with energy $\epsilon_3$, while the other is
absorbed at the horizon with energy $\epsilon_4<0$ (the reason for the
odd choice of subscripts will be clear soon). This results in
$\epsilon_3=\epsilon_0-\epsilon_4>\epsilon_0$, a net energy gain at
the expense of the rotational energy of the hole. Unfortunately this
process yields a modest efficiency $\epsilon_3/\epsilon_0 \leq
(1+\sqrt{2})/2\simeq 1.207$,
%
where the maximum is achieved for disintegration into two massless
particles~\cite{Bardeen:1972fi,Wald:1974kya,1975ApJ...196L.107P,Piran:1977dm}.

The original Penrose process relies on the somewhat unrealistic
disintegration of a single particle. An astrophysically more promising
variant of the idea is the {\em collisional Penrose
  process}~\cite{1975ApJ...196L.107P}: two bodies with energy
$(\epsilon_1,\,\epsilon_2)$ -- which could be elementary particles --
collide resulting in two bodies with energy
$(\epsilon_3,\,\epsilon_4)$.
This process has recently attracted interest in connection with dark
matter searches as a result of work by Banados, Silk and
West~\cite{Banados:2009pr}, who found that particle collisions near
rapidly rotating black holes could yield (in principle) unbound
center-of-mass energies, perhaps producing observable exotic ejecta
(see also \cite{Piran:1977dm}, \cite{Berti:2009bk,Jacobson:2009zg} for
caveats, and \cite{Harada:2014vka} for a review).
A crucial issue is that the highest center-of-mass energies are
achieved close to the horizon, and therefore the collision products
experience large gravitational redshifts, resulting in modest
efficiencies
\be
\eta\equiv \frac{\epsilon_3}{\epsilon_1+\epsilon_2}
\label{eta_def}
\ee
for the escaping particle (that we will assume, without loss of
generality, to be particle 3): $\eta\lesssim 1.5$, where the precise
upper bound depends on the nature of the colliding
particles~\cite{Piran:1977dm,Harada:2012ap,Bejger:2012yb}.

\begin{figure}[ht]
\begin{center}
\epsfig{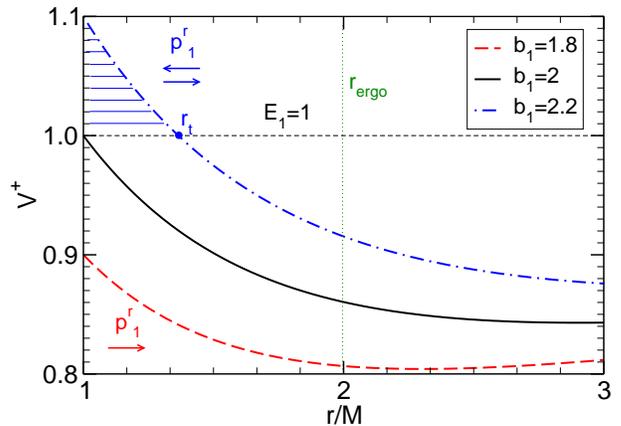}
\end{center}
\caption{\label{fig:effpot} Effective radial potential $V^+$
  [cf.~Eq.~\eqref{velocity}] for an extremal black hole, a particle
  with energy $E_1=1$ and angular momentum $L_1=b_1 E_1$. The horizon
  is located at $r=1$, and the ergoregion corresponds to $r<r_{\rm
    ergo}=2$. The dash-dotted (online: blue) curve shows the potential
  felt by particle 1 for the process discussed
  in~\cite{Schnittman:2014zsa}; particle 1 cannot access the dashed
  (blue) region with $r<r_t$.  The dashed (online: red) curve
  corresponds to the process considered in this paper. The solid
  (online: black) curve depicts the potential for the critical value
  of $b_1$ separating these two processes.}
\end{figure}

Surprisingly, Schnittman~\cite{Schnittman:2014zsa} recently reported
an order-of-magnitude increase in the efficiencies
($\eta \lesssim 15$) by allowing for a small kinematic change in the
collision, illustrated schematically in Fig.~\ref{fig:effpot}. One of
the colliding particles (say, particle 1) falls in the effective
potential for radial motion around an extremal black hole, defined in
Eq.~\eqref{velocity} below, and depicted by a dash-dotted (blue
online) curve in Fig.~\ref{fig:effpot}. Particle 1 rebounds at the
turning point $r=r_t$, so it has {\it outgoing} radial momentum
($p^r_1>0$) when it collides (at some radius $r_0$ such that
$r_t<r_0<r_{\rm ergo}$) with the incoming particle 2.  This outgoing
momentum favors ejection of a high-energy particle 3 after the
collision.

In this {\em Letter} we confirm the results
of~\cite{Schnittman:2014zsa}, and we reach the striking conclusion
that {\it arbitrarily large} efficiencies can be achieved when we do
not require both of the colliding particles to fall into the black
hole from infinity. In our ``super-Penrose'' processes the outgoing
particle 1 has angular momentum below the critical value for escape
(corresponding to the solid black line in Fig.~\ref{fig:effpot}) when
it collides with the ingoing particle 2. The radial motion of particle
1 is determined by the dashed (red online) effective potential:
particle 1 is confined to the vicinity of the black hole, and it must
have been created in the ergoregion via previous scattering events
(see \cite{Grib:2010xj} for a similar proposal). The likelihood of
such multiple scatterings is a delicate matter to be resolved by
detailed cross-section calculations for specific processes, but we
demonstrate that there is nothing preventing the creation of such
particles in the ergosphere (cf. the supplemental material): our
``super-Penrose'' collisions are {\em at least kinematically} allowed.
From now on we will use geometrical units ($G=c=1$).

\noindent{\bf{\em II. Setup.}}~Our setup is similar to that of
Ref.~\cite{Schnittman:2014zsa}. We consider two bodies (or particles)
1 and 2 colliding with four-momenta $p^{\mu}_1$ and $p^{\mu}_2$ at
Boyer-Lindquist coordinate position $r=r_0$ in the equatorial plane
($\theta=\pi/2$) of a Kerr black hole.
The final state consists of two bodies 3 and 4 with four-momenta
$p^{\mu}_3$ and $p^{\mu}_4$, also moving in the equatorial plane.  In
Boyer-Lindquist coordinates, the geodesic equations for equatorial
particles with rest mass $m$, energy $\epsilon\equiv
-g_{t\mu}p^{\mu}$ and angular momentum $\ell\equiv
g_{\varphi\mu}p^{\mu}$ (as seen by an observer at infinity) are given
by
\beq
\label{velocity}
\dot{r}^2&=&\frac{r^3+a^2(r+2M)}{r^3}\left(E-V^{+}\right)\left(E-V^{-}\right)\,,\nn\\
V^{\pm}&=&\frac{2aLM\pm \sqrt{r\Delta\left[L^2r+\left(r^3+a^2(r+2M)\right)\delta\right]}}{r^3+a^2(r+2M)}\,,\nn\\
\dot{\vartheta}&=&0\,,\nn\\
\dot{\varphi}&=&\frac{1}{\Delta}\left[\left(1-\frac{2M}{r}\right)L+\frac{2aM}{r}E\right]\,,
\eeq
%
where the dot denotes differentiation with respect to the geodesic
affine parameter $\lambda$, and $\Delta=r^2-2Mr+a^2$. Here $E\equiv
\epsilon/m$ and $L\equiv \ell/m$ denote the energy and angular
momentum per unit mass for massive particles, while for massless
particles this distinction does not apply ($E=\epsilon$ and
$L=\ell$). The constant $\delta=1~(0)$ for massive (massless)
particles, respectively. For massless particles the four-momentum is
$p^{\mu}=\dot{x}^{\mu}$; for massive particles we can
choose $\lambda=\tau/m$ ($\tau$ being proper time), so that
$p_{\mu}p^{\mu}=-m^2$. To enforce causality and guarantee that the
locally measured energy is always positive, physical geodesics must
satisfy $\dot{t}>0$.

%


Local conservation of four-momentum implies
\be
p^{\mu}_1+p^{\mu}_2=p^{\mu}_3+p^{\mu}_4\,.
\ee
Using Eqs.~\eqref{velocity}, this condition can be written as a set of
three equations:
\beq
\epsilon_1+\epsilon_2&=&\epsilon_3+\epsilon_4\,,\label{energy}\\
\ell_1+\ell_2&=&\ell_3+\ell_4\,,\label{angular}\\
p^r_1+p^r_2&=&p^r_3+p^r_4\label{radial}\,.
\eeq
The initial state is completely specified by fixing
$E_{1,2},\,L_{1,2}$ (and $m_{1,2}$ for massive particles). For massive
particles, we can specify the final state by providing a relation
between $m_3$ and $m_4$: for example, for two identical collision
products we can set $m_4=m_3$. Finally, we provide the Lorentz boost
$E_4$ and specific angular momentum $L_4$ and we are left with three
unknowns $(m_3,\,E_3,\,L_3)$ that can be determined by solving
Eqs.~\eqref{energy}--\eqref{radial}. To be observationally
interesting, the solutions of the system above must satisfy the
requirement that particle 3 can escape and reach an observer at
infinity (i.e., there cannot be any turning points $r^{\rm t}_{3}$ for
particle 3 with $r^{\rm t}_{3}>r_0$). To find the maximum efficiency
$\eta_{\rm max}$, we simply repeat the previous procedure for a range
of values of $L_4$. We find that the efficiency only depends on the
impact parameters of the initial particles
$b_{1,2}\equiv L_{1,2}/E_{1,2}$ and on the ratio of their energies
$R\equiv \epsilon_1/\epsilon_2$. Therefore all of our results will be
shown as functions of these quantities. For simplicity, we will also
set the black-hole mass $M=1$.

\begin{figure}[th]
\begin{center}
\epsfig{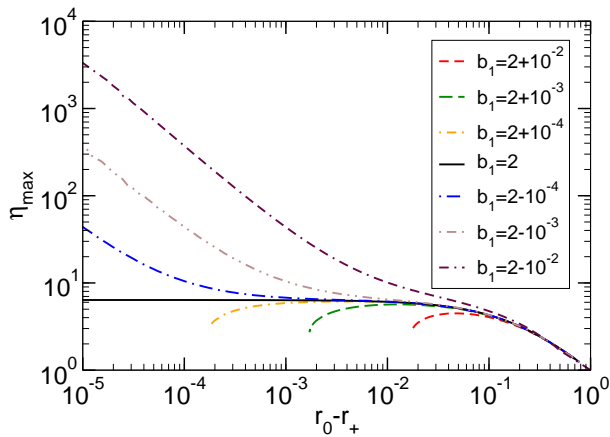}
\end{center}
\caption{\label{fig:penrose_schnittman} Maximum efficiency $\eta_{\rm
    max}$ for the collision of equal-energy particles ($R=1$) as a
  function of the radius at which the reaction occurs. We consider
  $p_1^r>0$, $p_2^r<0$, $b_2=-2(1+\sqrt{2})$, and an extremal black
  hole ($a=1$). The curves for $b_1>2$ terminate at the turning point
  of particle 1.}
\end{figure}
\begin{figure}[ht]
\begin{center}
\epsfig{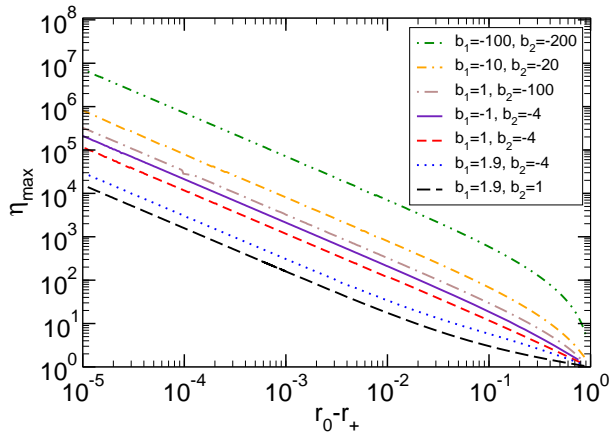}
\end{center}
\caption{\label{fig:penrose_huge} Maximum efficiency $\eta_{\rm max}$
  for the same process considered in Fig.~\ref{fig:penrose_schnittman}
  as a function of the radius at which the reaction occurs, for
  $a=1$. When $b_{2}<-2(1+\sqrt{2})$, particle 2 must also be produced
  close to the black hole.}
\end{figure}

\begin{figure}[th]
\begin{center}
\epsfig{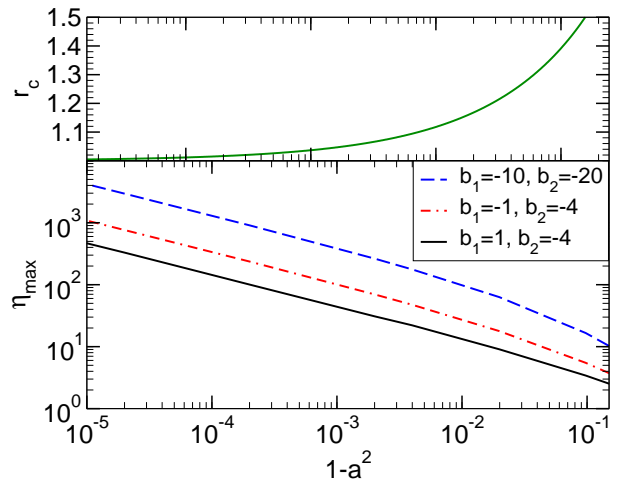}
\end{center}
\caption{\label{fig:penrose_vs_a} Maximum efficiency as a function of
  the black hole spin $a$ for the same process as in
  Fig.~\ref{fig:penrose_schnittman}. The maximum efficiency is
  attained at the photon sphere, $r_{c}$, below which the infalling
  particle 3 can not escape. For $a\simeq 1$, the maximum efficiency
  scales as $1/\sqrt{1-a^2}$.}
\end{figure}

\begin{figure}[ht]
\begin{center}
\epsfig{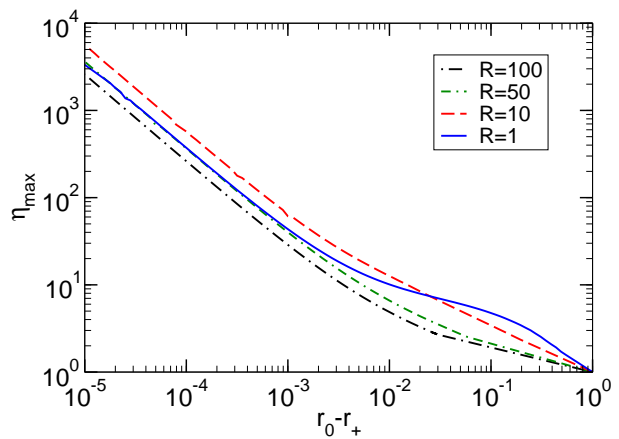}
\end{center}
\caption{Maximum gain for a Compton-like scattering between a massless
  particle with $p_1^r>0$, $b_1=1.99$ and a massive particle with
  $p_2^r<0$, $b_2=-2(1+\sqrt{2})$. We plot the maximum gain for
  different values of
  $R\equiv \epsilon_1/\epsilon_2$.\label{fig:compton}}
\end{figure}

\noindent{\bf{\em III. Super-amplification collisions.}}~We focus on
collisions for which particle 1 is outgoing (i.e., $p^{r}_1>0$) while
particle 2 is ingoing. Our results are summarized in
Figs.~\ref{fig:penrose_schnittman}--\ref{fig:compton}. We start by
using the same initial conditions as in~\cite{Schnittman:2014zsa}: an
extremal black hole and $E_1=E_2$. For $b_1=2$ we confirm that in this
case the peak efficiency is $\eta_{\rm max}\sim 6.4$ (a value sensibly
larger than the predictions of previous studies
\cite{Harada:2012ap,Bejger:2012yb,Harada:2014vka}) for collisions that
occur close enough to the horizon. In
Fig.~\ref{fig:penrose_schnittman} we show the maximum efficiency,
varying $b_1$ close to the critical value $b_1=2$ and fixing the
impact parameter of particle 2 to the value $b_2=-2(1+\sqrt{2})$.
We find that the maximum efficiency quickly decreases with increasing
$b_1$, because particle 1 is now allowed to move in a smaller region
inside the ergosphere (see Fig.\ref{fig:effpot}). For sub-critical
particles ($b_1<2$) the scenario is completely different, and the
efficiency becomes arbitrarily large as we approach the horizon. In
our searches we found that these ``super-Penrose collisions'' occur
for $b_2<b_1<2$.

Fig.~\ref{fig:penrose_huge} shows that the efficiencies of
super-Penrose collisions can be orders of magnitude larger than those
found in \cite{Schnittman:2014zsa} in a large region of the parameter
space of initial conditions.  For very large $(|b_1|,\,|b_2|)$ and
collisions close to $r=r_+$, the peak efficiency scales as $\eta_{\rm
  max}\sim 0.5\sqrt{(2-b_1)(2-b_2)}/(r_0-r_+)$.
Efficiencies $\eta\gtrsim 10^3$ are easily achieved for maximally
spinning black holes.


Fig.~\ref{fig:penrose_vs_a} shows the dependence of the
super-amplification efficiency on the black-hole spin $a$. For $a<1$
the maximum gain quickly drops (bottom panel), and particle 3 can only
escape for collisions occurring at $r_0>r_c$, where $r_c$ (top panel)
is the Boyer-Lindquist radius of the photon sphere in the equatorial
plane. The maximum efficiency is attained when the collision occurs
precisely at the photon sphere, where we find the scaling
$\eta_{\rm max}\sim (1-a^2)^{-1/2}$. Nonetheless, very high
efficiencies are still allowed for astrophysically relevant black
holes, as long as particles with large negative impact parameters
$b_1$ and $b_2$ are created inside the ergosphere and collide. For
collisions where particle 2 falls from spatial infinity and black
holes spinning at the Thorne limit $a\sim 0.998$ \cite{Thorne:1974ve}
we still find efficiencies as large as $\eta \sim 10^2$.

Finally, in Fig.~\ref{fig:compton} we consider the Compton-like
scattering of a massless particle with $p_1^r>0$ and a massive particle with $p_2^r<0$ and we show the
dependence of the maximum efficiency on the energy ratio $R\equiv \epsilon_1/\epsilon_2$ of the
initial particles. For reactions occurring near the edge of the
ergoregion the efficiency is largest when $R=1$, while near the
horizon ($r/r_+-1\ll 1$) the maximum efficiency is achieved for
$R\approx 10$. By comparing the solid (blue online) and
dash-dash-dotted (green online) curves, we see that the maximum
efficiency at the horizon occurs for $1<R\lesssim 50$, and for
$R\gtrsim 50$ the efficiency decreases again. In conclusion, reactions
such as the inverse Compton scattering between photons and massive
particles could (at least in principle) produce highly energetic
photons by multiple scattering processes.

\noindent{\bf{\em IV. The origin of the colliding
    particles.}}~Super-Penrose amplification cannot result from the
collision of two particles coming from spatial infinity: as shown
schematically in Fig.~\ref{fig:effpot}, particle 1 (with $p_1^r>0$ and
$b_1<2$ for $a=1$) must be created inside the ergosphere by previous
scattering events. Furthermore, for $b_{1,2}<-2(1+\sqrt{2})$ there are
turning points of the motion outside the ergoregion, and particles
would be deflected back before reaching the ergoregion (cf. Fig.~1
of~\cite{Schnittman:2014zsa}).

As argued at length in the supplemental material, the colliding
particles giving rise to super-amplification are physically relevant
initial states, because they can be created (for example) by the
collision of particles coming from infinity. In fact there is a wide
range of realistic initial conditions that can result in
``super-Penrose initial conditions''. For example, a particle with
$E_1=1,\,m_1=1,\,L_1=1.9$ (dotted blue line in
Fig.~\ref{fig:penrose_huge}) in the extremal Kerr background can be
generated by two particles with rest masses $(1,m_*)$ falling from
rest at infinity
%
%
with angular momenta $(L_1=-4,L_2=2)$ and colliding at $r=1.01$, as long as $m_*>6$.
The threshold mass ratio $m_*$ depends on the angular momentum of the
colliding particles and on the spin of the black hole: it is
proportional to $|L_1|$, and it scales like $(r-1)^{-2}$ for extremal
black holes.
Note in particular that we can have super-Penrose collisions for
arbitrarily large negative angular momenta of particle 2. Due to frame
dragging, the cross section for counterrotating incoming particles is
much larger. Counterrotation may be a common feature in astrophysics,
e.g. because of disk fragmentation, as in the ``chaotic accretion''
scenario~\cite{King:2006uu,Berti:2008af}.
Finally, we could (very conservatively) define a combined efficiency
$\eta'$ so that $\epsilon_{1,2}$ in Eq.~\eqref{eta_def} refers to the
``parent'' particles falling from infinity. With this redefinition,
the maximum efficiency typically seems to lower to the levels
predicted in~\cite{Schnittman:2014zsa} (cf. the supplemental material
and Sec.~V below).

In summary, the initial conditions giving rise to super-amplification
are kinematically allowed as the result of collisions of particles
falling into the hole from large distances. 

\noindent{\bf{\em V. Multiple scattering.}}~A consequence of the
argument presented above is that multiple scattering events can
increase the energetic gain achievable with (and the astrophysical
relevance of) the ``traditional'' collisional Penrose process
\cite{1975ApJ...196L.107P,Piran:1977dm,Banados:2009pr} and of
Schnittman's variant of the process \cite{Schnittman:2014zsa}. This is
because the energy of particles that cannot escape to infinity may be
substantially larger than the energy of those that can. Even if
``trapped'' and unable to escape themselves, these particles may
collide with other particles and give rise to high-energy collision
products that {\em may} escape and be detectable.

Multiple energy-extracting collisions may lead to very large
efficiencies~\cite{Schnittman:2014zsa}. Each Penrose scattering
decreases the black-hole spin, but efficiencies can still be
moderately high, even away from $a=1$. These events may well be rare,
but it is tempting to propose that they could play a role in the
production of observable gamma rays or ultra-high-energy cosmic
rays. A detailed assessment of these possibilities is beyond the scope
of this {\em Letter}.

\noindent{\bf{\em VI. Discussion.}}~Previous
studies~\cite{Schnittman:2014zsa,Harada:2012ap,Piran:1977dm,Bejger:2012yb}
(which we reproduced) focused on a region of the parameter space that
excludes by construction the amplification mechanism studied in our
work. The astrophysical likelihood of the super-Penrose amplification
process proposed here is obviously a critical issue that requires
further work. For very large efficiencies, the energy of the escaping
particle can be as large as the black hole's, and our geodesic
approximation clearly breaks down. In this regime back-reaction
effects must be included.  Furthermore, numerical simulations and
detailed calculations of production rates (along the lines
of~\cite{Piran:1977dm,Banados:2010kn,Williams:2011uz}) are needed to
make conclusive statements about the astrophysical relevance of these
results.

We made the simplifying assumption that the reaction occurs in the
equatorial plane. Very few off-equatorial calculations of the Penrose
process have been performed in the past, mainly due to computational
difficulties. The results reported here, together with recent studies
of the off-equatorial collisional Penrose
process~\cite{Harada:2011xz,Gariel:2014ara}, suggest that surprises
may be in store, and a generalization of our calculations to the
off-equatorial case is urgently needed.

Another important extension of our work would be the inclusion of external magnetic fields, a common feature of astrophysical black holes.  The
Penrose process for charged particles in the presence of
electromagnetic fields is known to be more efficient than the original
process~\cite{Wagh:1986tsa,1986ApJ...301.1018W}, with efficiencies as
large as $\eta\sim 10$~\cite{1986ApJ...307...38P}.  The possibility
that electromagnetic fields could trigger and enhance super-Penrose
collisions makes this an important line of research for the future.

We hope that this {\em Letter} will stimulate further work in these
directions and improve our understanding of some of the most energetic
events in the Universe.

After the completion of this paper we learned about work by
Zaslavskii~\cite{Zaslavskii:2014jea} that confirms our main findings.

\noindent{\bf{\em Acknowledgments.}}
We thank Jorge Rocha and Jeremy Schnittman for useful comments.
E.B. was supported by NSF CAREER Grant No.~PHY-1055103.
R.B. acknowledges financial support from the FCT-IDPASC program
through the Grant No. SFRH/BD/52047/2012, and from the Funda\c c\~ao
Calouste Gulbenkian through the Programa Gulbenkian de Est\' imulo \`a
Investiga\c c\~ao Cient\'ifica.
V.C. acknowledges financial support provided under the European
Union's FP7 ERC Starting Grant ``The dynamics of black holes: testing
the limits of Einstein's theory'' grant agreement no. DyBHo--256667.
This research was supported in part by the Perimeter Institute for
Theoretical Physics.  Research at Perimeter Institute is supported by
the Government of Canada through Industry Canada and by the Province
of Ontario through the Ministry of Economic Development $\&$
Innovation.
This work was supported by the NRHEP 295189 FP7-PEOPLE-2011-IRSES
Grant.
%

%

\clearpage

\noindent{\bf{\em Supplemental Material.}}~After our work was
submitted, Leiderschneider and Piran~\cite{Leiderschneider:2015ika}
posted a comment to the arXiv where they argue that the process
discussed in our {\em Letter}, while kinematically allowed, ``requires
a deposition of energy that is divergently large compared with the
energy of the escaping particle.'' They conclude that the efficiency
of the combined process is extremely small. They also state that the
total energy gain of the collisional Penrose process, even accounting
for our new findings, must be ``modest.''
The calculations of~\cite{Leiderschneider:2015ika} are a special case
of our own and we confirmed independently their Eq.~(4), but we
disagree with both conclusions.

In Section IV of our initial arXiv submission we presented a {\em
  specific} example in which a particle with positive radial momentum
(necessary for our ``super-Penrose'' collisions) can emerge in the
ergosphere. The example consisted of two particles falling with zero
angular momentum from rest at infinity.  We pointed out that these two
particles produce a particle in the ergoregion with the required
properties (positive radial momentum and potentially divergent
amplification factors) when their mass ratio $m_*$ is large enough
(roughly, $m_*>10^3$). The critical mass ratio depends on how close to
the horizon the particles collide, and it diverges as we approach the
horizon.

Ref.~\cite{Leiderschneider:2015ika} focused exclusively on this
example. It argued that we should not use the traditional definition
of efficiency, but rather consider the {\em combined} efficiency of
the two collisions:
\be\label{etaprime}
\eta'=\frac{\epsilon_3}{\epsilon'_1+\epsilon'_2+\epsilon_2}\,,
\ee
where the primed quantities $\epsilon'_{1}=m'_1=1$ and
$\epsilon'_{2}=m'_2=m_*$ are the energies of the ``parent'' particles
falling from rest at infinity to produce particle 1.

\begin{figure*}[thb]
  \includegraphics[width=0.9\columnwidth,clip=true]{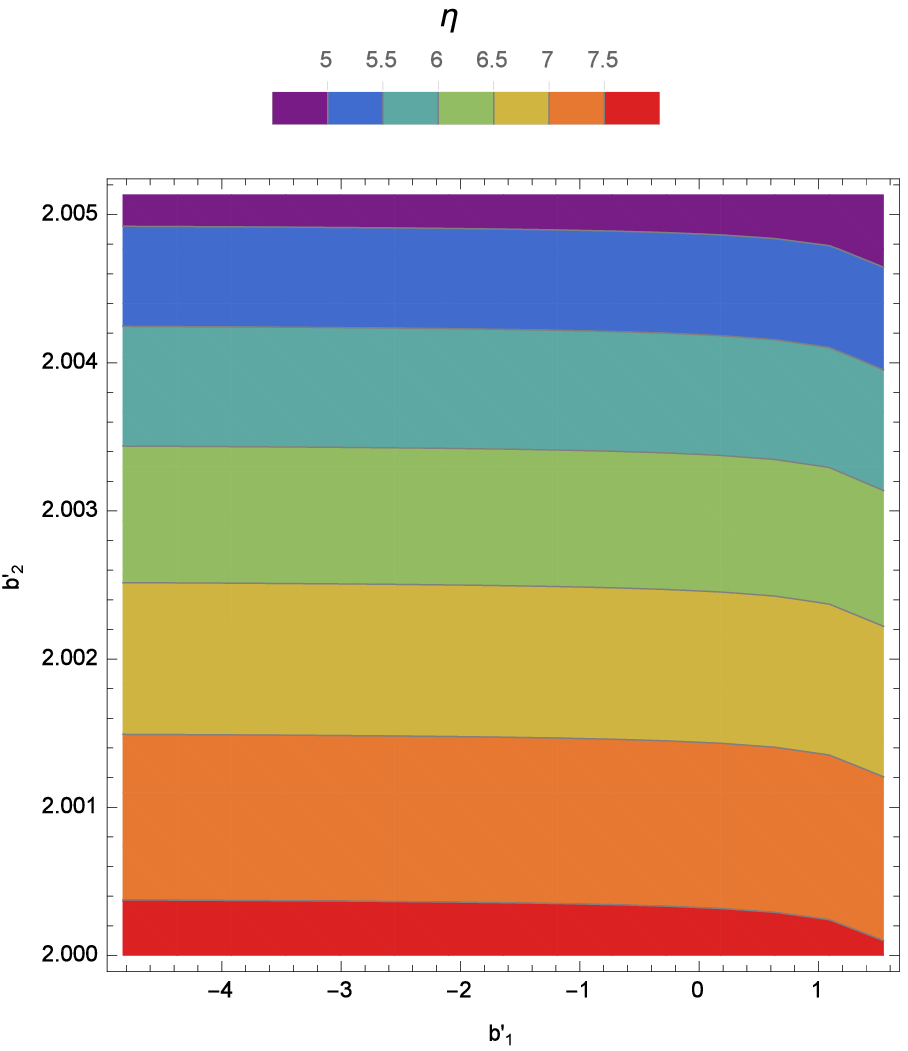}
  \includegraphics[width=0.9\columnwidth,clip=true]{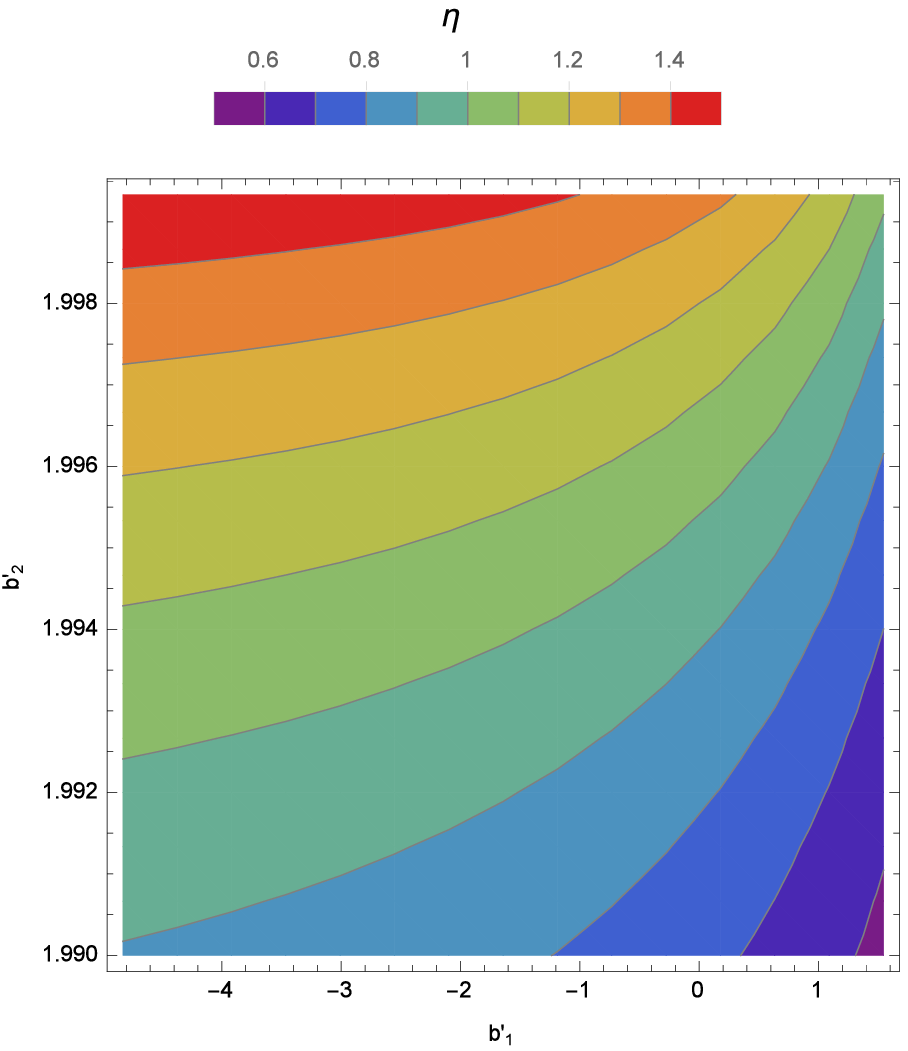}\\
  \caption{Left panels: Contour plots within a range of impact
    parameters $(b'_1,b'_2)$ of the combined efficiency $\eta'$
    defined in Eq.~\eqref{etaprime}, where $\epsilon'_{1}=E'_1 m'_1=1$
    and $\epsilon'_{2}=E'_2 m'_2=m_*$ are the energies of the
    particles falling from rest ($E'_1=E'_2=1$) at infinity. Here
    particle $2'$ has a positive radial momentum $p^{r}_{2'}>0$ at the
    point of collision, which takes place at $r=1.01$. The {\em
      combined} efficiencies defined in this way are of the order of
    what was found by Schnittman~\cite{Schnittman:2014zsa}, at odds
    with the claims of~\cite{Leiderschneider:2015ika}. Right panels:
    same but for the case where particle $2'$ has negative radial
    momentum $p^{r}_{2'}<0$ at the point of collision. With this
    restriction, the {\em combined} efficiencies reduce further to the
    ``pre-Schnittman'' levels
    of~\cite{Bejger:2012yb}. \label{fig:etap_r101}}
\end{figure*}

\begin{figure*}[thb]
  \includegraphics[width=0.9\columnwidth,clip=true]{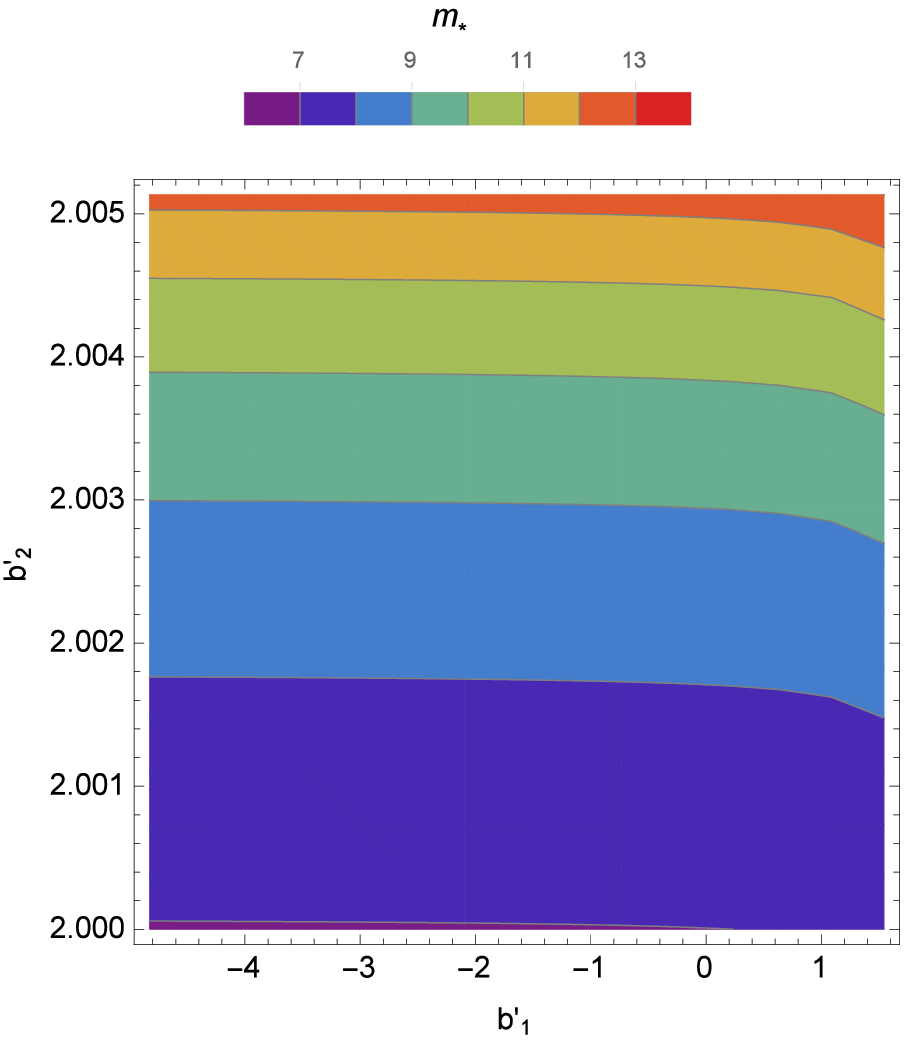}
  \includegraphics[width=0.9\columnwidth,clip=true]{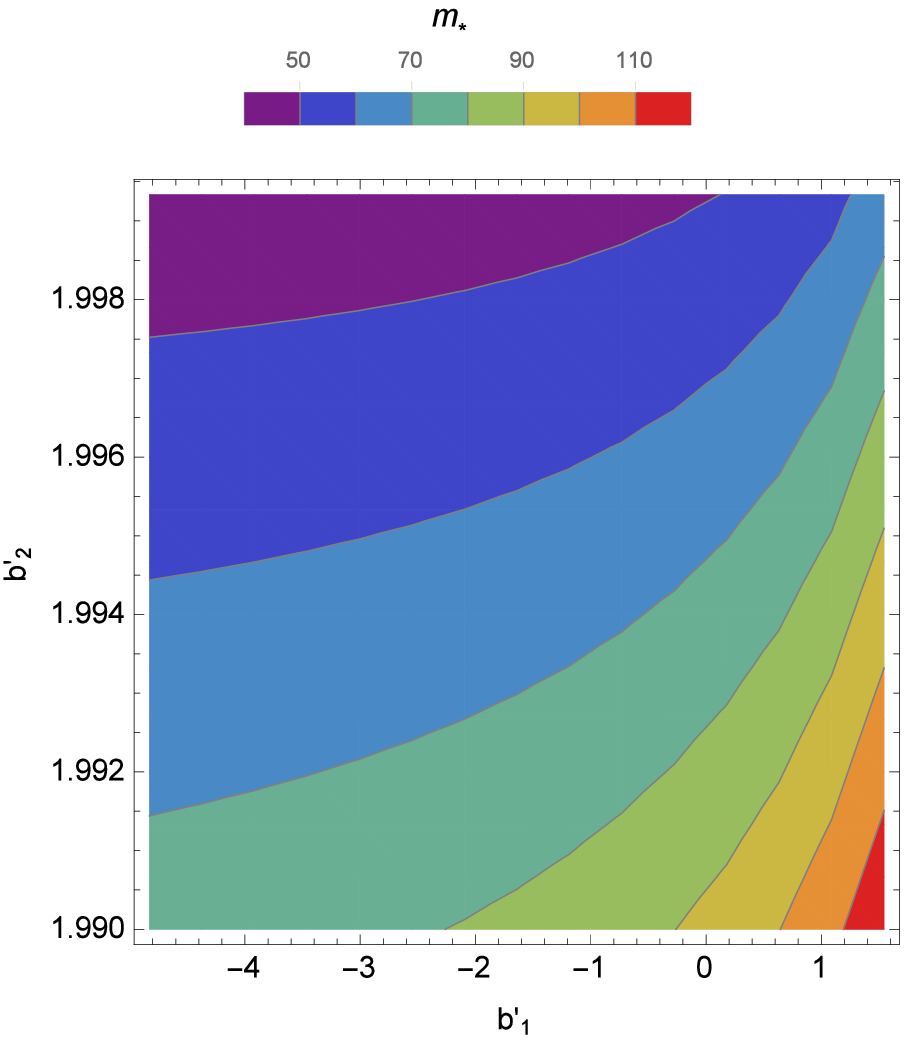}
  \caption{Left panels: Contour plots within a range of angular
    momenta $(b'_1,b'_2)$ of the mass ratio $m_*$ giving the
    efficiencies plotted in Fig.~\ref{fig:etap_r101} when particle $2'$
    has a positive radial momentum $p^{r}_{2'}>0$ at the point of
    collision, which occurs at $r=1.01$. Right panels: same but when
    particle $2'$ has negative radial momentum $p^{r}_{2'}<0$ at the point
    of collision. \label{fig:mstar_r101}}
\end{figure*}

We disagree with this unnecessary assumption, that restricts the
``initial conditions'' leading to collisions in the ergosphere in a
way that is unnatural in astrophysics (see below). However, even under
this more restrictive definition, the conclusion
of~\cite{Leiderschneider:2015ika} that ``the super-Penrose collision
results in a net (extremely large) energy loss'' is wrong, as we show
here by working out a more generic case. We consider two particles
falling from rest at infinity within a range of impact parameters
$(b'_1,\,b'_2)$. We let them collide at $r=1.01$ (for concreteness)
and we draw contour plots of the combined efficiency $\eta'$
(Fig.~\ref{fig:etap_r101}) and of the mass ratio $m_*$ necessary to
produce the given $\eta'$ (Fig.~\ref{fig:mstar_r101}).
These contour plots address the concern
of~\cite{Leiderschneider:2015ika}: there {\em is} clearly an energy
gain in this process. This energy gain occurs {\em generically}, and
the required $m_*$ are relatively small.
The combined efficiencies are of the same order as those found
in~\cite{Schnittman:2014zsa} (and sometimes a little larger) when one
of the parent particles is allowed to have positive radial momentum,
and of the same order as those in~\cite{Bejger:2012yb} when both
parent particles have negative radial momentum (as in all work prior
to~\cite{Schnittman:2014zsa}).
By allowing for subsequent scattering the efficiency can be made
higher (and it can become unbound for the unrealistic case with
infinite scatterings).

In conclusion: combined efficiencies are {\em not} ``less than unity''
as claimed in the conclusions of~\cite{Leiderschneider:2015ika}: this
is only a consequence of the small parameter space (parent particles
with zero angular momenta) considered in their work. Of course the
{\em combined} amplification is not super-Penrose, because (by
construction!) we are now limiting the parameter space to that of
Ref.~\cite{Schnittman:2014zsa} (in our left panels) and to that before
Schnittman's work (in our right panels).
This result is expected, as we point out at the end of Section IV of
our submission, where we observe that ``with this redefinition, the
maximum efficiency typically seems to lower to the levels predicted
in~\cite{Schnittman:2014zsa}.''



Crucially, this does not mean that larger amplifications cannot occur
in nature: it only means that we are considering {\em a very small
  kinematical subset of all collisions} -- those where ``our''
particle 1 results from particles falling from rest at infinity.
%
In Section IV we gave a specific, concrete example (inspired by
Penrose's original idea) of particle collisions leading to ``initial
data'' for a super-Schnittman amplification condition. For simplicity,
we took the simplest case: that of particles falling radially from
rest. This is the only reason why we found a large $m_*$. If particles
fall with generic angular momenta from infinity, $m_*$ is sensibly
lower. We {\em still} get combined efficiencies $\eta'>1$, but they
are now reduced to Schnittman-like levels. The point of our {\em
  Letter} is that {\em requiring this is highly restrictive}. This is
analogous to how considering only collisions between particles with
negative radial momenta before the collision was limiting efficiencies
in all papers before Ref.~\cite{Schnittman:2014zsa}.

The main novelty of our results with respect to those of
Ref.~\cite{Schnittman:2014zsa} is precisely that when we allow
particle 1 to be produced in the ergoregion by colliding particles
that do {\em not necessarily} come from infinity, we are {\em sensibly
  raising the likelihood} of a very efficient collisional Penrose
processes.
In a realistic astrophysical environment, the particles colliding in
the ergoregion will typically fall in from (say) an accretion disk,
and not as a result of idealized collisions between particles falling
from infinity. The kinematical conditions under which these collisions
occur, and the likelihood that they result in super-Schnittman
amplification, can only be assessed by Monte Carlo simulations. This
is beyond the proof-of-principle scope of our {\em Letter}.

In astrophysical environments, all sorts of interactions can occur
that produce particles with positive radial momentum {\em inside} the
ergosphere. Of course particle production via collisions inside the
ergosphere is nothing exotic: it is the essence of the original
(noncollisional) Penrose process. In our opinion, our finding is
rather striking: whenever such a particle exists, it can generate
efficiencies that are orders of magnitude larger than those
in~\cite{Schnittman:2014zsa}.
More work is definitely needed to establish whether this idea is
likely to explain large energy outflows around rotating black holes,
and we hope that our {\em Letter} will stimulate precisely this kind
of work.


\end{document}